\documentclass[english,twocolumn,pra,aps,showpacs]{revtex4}

\usepackage{amsfonts}
\usepackage{bbold}
\usepackage[dvips]{graphicx}
\usepackage{babel,amsmath,amssymb,float}
\usepackage{psfrag}
\usepackage{pstricks}
\usepackage{pst-node}
\usepackage{pst-plot}

\newcommand{\qed}{\hspace*{\fill}$\square$}

\newcommand{\be}{\begin{equation}}
\newcommand{\ee}{\end{equation}}

\begin{document}

\title[Short Title]{Google in a Quantum Network}

\author{G.D. Paparo and M.A. Martin-Delgado}
\affiliation{ Departamento de F\'{\i}sica Te\'orica I, Universidad
Complutense, 28040. Madrid, Spain. }

\begin{abstract}
We introduce the characterization of a class of quantum PageRank algorithms in a scenario in which
some kind of quantum network is realizable out of the current classical internet web, but no quantum computer is yet available.
This class represents a quantization of the PageRank protocol currently employed to list web pages according to their importance.
We have found an instance of this class of quantum protocols that outperforms its classical counterpart and may break the classical
hierarchy of web pages depending on the topology of the web.
\end{abstract}


\maketitle

\section{Introduction}
\label{sect:intro}

The possibility of establishing a quantum network of practical use is currently
under active investigation. Some early versions of them, modest as they may be, have been
designed and realized in real world in the recent years \cite{darpa,SECOQC,UQCC,SwissQuantum,vicente1,ETSI}, 
or in some instances they are under way.
In fact, building a quantum network has been targeted as a fundamental goal in quantum information \cite{NC,rmp}
an even a more feasible goal to accomplish than the first scalable quantum computer. 
There are other more advanced proposals for quantum networks \cite{quantum_internet1,quantum_internet2} based on entanglement connections
that need quantum repeaters \cite{qrepeaters98,qrepeaters99,rmprepaters} in order to function properly and being stable 
\cite{telecomunications11,qap10,telecomunications10}. Related to this, some physical quantum models exhibit very remarkable long-distance
entanglement properties \cite{mps03,le04,korepin,graph_states}.
Another alternative to build different types of quantum networks makes use of quantum percolation protocols \cite{percolation07,distribution10,calsamiglia09,calsamiglia11}.

Thus, it is interesting to study how different possibilities of  quantum networks would behave regarding what we know about the world wide web. 
In particular, an essential ingredient in the classical network that we enjoy today is the ability to search web pages in the immense changing world that the web has come to be 
known. The key tool for performing those searches is the notion of  the PageRank algorithm 
\cite{BrinPage98,BrinPage98a,BrinPage98b,LangvilleMeyer04,Meyer00,HaveliwalaKamvar03,marijuan11,shepelyansky,vitanyi}.

We envision a similar tool to perform the task of ranking quantum webpages. A quantum webpage is a node of the quantum network realized by means of a quantum storage device, like a quantum memory. What is crucially important is that a fully fledged quantum computer is not required. Thus, the quantum webpages have limited capabilities:  storing and reading in/out quantum states.

Since the notion of a quantum PageRank is by no means unique, it is convenient to introduce a class or category of possible quantum PageRanks.
They must satisfy a set of properties that define an admissible class:

{\bf Quantum Computer PageRank Class}:
\noindent \begin{description}

\item[Q1] The classical PageRank must be embedded into the quantum class
in such a way that the directed graph structure is preserved at the quantum level.

\item[Q2] The sum of all quantum PageRanks must add to 1 i.e. $\sum_i I_q(P_i) = 1$.

\item[Q3] The Q-PageRank admits  a quantized Markov Chain (MC) description. 

\item [Q4] The classical algorithm to compute the quantum computer PageRank belongs to the computational complexity 
class BQP. 

\end{description}

However, as we have noticed that working with a real quantum computer seems not realistic in the near future and prospects for 
realizing quantum networks based on nodes made up of quantum storing devices, like quantum memories, are high as shown by
experimental progress, we find also very convenient to introduce a more realistic class of quantum PageRanks as follows:

{\bf Quantum PageRank Class}:
\noindent \begin{description}

\item[P1] The classical PageRank must be embedded into the quantum class
in such a way that the directed graph structure is preserved at the quantum level.

\item[P2] The sum of all quantum PageRanks must add to 1 i.e. $\sum_i I_q(P_i) = 1$.

\item[P3] The Q-PageRank admits  a quantized Markov Chain (MC) description. 

\item [P4] The classical algorithm to compute the quantum PageRank belongs to the computational complexity 
class P. 

\end{description}

The crucial difference is the substitution of property Q4 that deals with the feasibility of calculations with a quantum computer in polynomial time
by property P4 which is more demanding since it means that even a classical computer can simulate the process and find the importance of 
webpages in polynomial time.

Property P1 reflects the fact that originally we would have an internet that is a classical network represented by a 
directed graph and then shall apply some kind of quantization procedure in order to turn it into a quantum network.
The latter must be compatible with the classical one, particularly preserving the directed structure which is crucial to measure a page's  authority. This is non-trivial
and some quantization methods may fail to produce a unitary quantum PageRank importance for the quantum case,
as shown in Sect.\ref{sect_III}.

With property P2 we guarantee that we have a globally well-defined notion of the importance of a web page at the quantum level.
This allows us to have the probabilistic interpretation of the surfer's position (see Sect.\ref{sect_III}).

Property P3 is the key to a wide class of natural quantization methods for the classical PageRank based in the equivalence of this one
with a classical Markov chain process (see Sect.\ref{sect_III}). Thus, it is natural that the equivalent property holds true in the quantum version of the PageRank,
and consequently, its description in terms of a quantized surfer's motion.

The reason for property P4 relies on the assumption that we envisage a near-future scenario when a certain class of quantum network will be operative
but not yet a scalable quantum computer. Therefore, we demand that the computation of the quantum PageRank $I_q$ be efficiently carried out
on a classical computer.

In this paper we have constructed a valid quantum PageRank that fulfills all these requirements. We remark that there may be other solutions to the quantum version
of the PageRank within the class defined above, but nevertheless we shall show that finding one instance of this quantum PageRank class is a non-trivial task.

The definition of Class of Quantum PageRank given here  is very general and it can accommodate very diverse situations,
ranging from quantum channels (including entangled states) between nodes in the network to simpler situations where network nodes are quantum levels
of a multilevel quantum system where information is stored in the same way as in the Grover algorithm. 
In the particular instance of Quantum PageRank algorithm that we have found in subsection  {\it Results: A Quantization of Google PageRank} , we are concerned with the latter case of a multilevel quantum
systems where the quantum state is attached to a link of the network of nodes. This instance does not exhaust all possibilities represented in our generic Class
of Quantum PageRanks by all means, but it is the first non-trivial instance in which sharp deviations from the classical PageRank algorithm can be found,
as explained later on in our paper.

The explicit step-by-step description of our quantum PageRank algorithm is presented in Sect.\ref{sect_IV}. A key distinctive feature of this quantum algorithm is that
the importance of the quantum pages exhibit quantum fluctuations unlike its classical counterpart. These quantum fluctuations, as shown in the simulations in Sect.\ref{sect_V},
show up in the form of time dependent importances $I_q(P_i,t)$, which causes in turn that sometimes one certain pair of pages satisfy $I_q(P_i,t_1)>I_q(P_j,t_1)$, 
and some other times the relative importance is reversed $I_q(P_i,t_2)<I_q(P_j,t_2)$, for time-steps such that  $t_1<t_2$.
We may use an analogy to understand this situation: the classical PageRank gives us a snapshot or photo with a fixed hierarchy of web pages according to their calculated
importance. On the contrary, the quantum PageRank is more like a movie since the quantum importance of the pages vary with time. In order to produce a fixed output
made of a list with the quantum pages sorted according to their importance, a natural choice we make is to compute the temporal average of the quantum PageRanks
and their standard mean deviation.

There are additional features that a certain quantum PageRank may have as a consequence of the definition of the class above.
We provide hereby several useful definitions:

\noindent {\bf  Strong Hierarchical Preserving PageRank}: when  the classical hierarchical structure of a PageRank  is preserved upon quantization.

\noindent This notion is too strong when the PageRank varies with time as it is the case at the quantum level.

\noindent {\bf  Weak Hierarchical Preserving PageRank}: when the node with highest classical PageRank is preserved after quantization,  but not so for the rest of pages.

\noindent {\bf   Outperforming}: when the highest classical PageRank of a page is overcome by the quantum PageRank of that page.

Outperforming may occur at one given instant or on average thereby leading to the natural extended concepts of  instantaneously outperforming or average outperforming,
respectively.

In section \ref{sect_conclusions}  we provide a list of main results that we have obtained with our quantized version
of the quantum PageRank algorithm. Remarkably, quantum fluctuations may change the classical hierarchy of web pages
both instantaneously and in terms of mean values.

This paper is organized as follows:
in Sect.\ref{sect_II} we give an introduction to the
classical notion of PageRank needed to present
in Sect.\ref{sect_III} the quantum version of it.
In Sect.\ref{sect_IV}  we present our proposal for a quantum version 
of Google PageRank algorithm.
In Sect.\ref{sect_V} we perform numerical simulations of the 
quantum algorithm to representative directed graphs representing
either an intranet or a general web with no special symmetry.
Sect.\ref{sect_conclusions} is devoted to conclusions.

\section{Classical PageRank}
\label{sect_II}

Brin and Page introduced Google in 1998~\cite{BrinPage98,BrinPage98a,BrinPage98b}, a time when the pace  at which the web was growing began to outstrip the ability of current search engines to yield useable results. A major distinction between their algorithm, called PageRank (PR),  and previous approaches is the fact that PR has an objective character, while other searchers  were based up on the subjective criterium of the contents of the pages , because they were built as a collection of links that people in companies  stored on a regular basis. 
In order words, PR is dynamical while the other approaches were static and subjective w.r.t. contents of the pages. 

 The way most search engines, including Google, work is to continually retrieve pages from the web, index the words in each document, and store this information. Each time a user asks for a web search using a search phrase, such as "search engine", the search engine determines outputs  all pages on the web that contain the words in the searched phrase or are semantically related to it. 
 
 Then, a problem arises naturally: Google now claims to index 50  billion pages. Roughly 95 \% of the text in web pages is composed from a mere 10,000 words. This means that, for most searches, there will be a huge number of pages containing the words in the search phrase. What is needed is a mean of ranking the importance of the pages that fit the search criteria so that the pages can be sorted with the most important pages at the top of the list. Their success is largely due to PageRank's ability to rank the importance of pages in the WWW.

\subsection{Google PageRank}

The key idea of Google's PageRank algorithm is that the importance of a page is given by how many pages link to it. 
If we define $I(P_i)$ as the importance of a page $P_i$ and $B_i$ as the set of pages linking to it, then we might think to put in equations the key idea put forward above as follows:
\begin{equation}
I(P_i):= \sum_{j \in B_i} \frac{I(P_j)}{\rm{outdeg}(P_j)},
\label{ eq: definition of pagerank1}
\end{equation}
where outdeg($P_j$) is the outdegree (i.e. the number of outgoing links) of the page $P_j$.
Let us define a matrix, called the hyperlink matrix, in which the entry in the ith row and jth column is:
\begin{equation}
H_{ij} := 
\begin{cases} 
\;1/\rm{outdeg}(P_j)  & \mbox{if }P_j \in B_i  \\
\;0 & \mbox{otherwise }
\end{cases}
\label{eq:definition hyperlik matrix}
\end{equation}
We will also form a vector $I$ whose components are the PageRanks $I(P_i)$. 
The condition above defining the PageRank can be expressed in matrix form as:
\begin{equation}
I = H I 
\label{ eq: definition of pagerank1 matrix form}
\end{equation}
Thus, we have recast  the problem of finding the PageRanks as the problem of finding the eigenvalues of a matrix~\cite{LangvilleMeyer04}. We are in for a special challenge since the matrix H is a square matrix with one column for each web page indexed by Google. This means that H has about  $n=50$  billion columns and rows. However $H$ is a sparse matrix, i.e. most of the entries in H are zero; in fact, studies show that web pages have an average of about 10 links, meaning that, on average, all but 10 entries in every column are zero. 

We will choose a method known as the power method for finding the stationary vector I of the matrix H. How does the power method work? We begin by choosing a vector $I_0$ as a candidate for $I$ and then producing a sequence of vectors $I_k$ by:
\begin{equation}
I^{k+1} = H I^{k} 
\label{ eq: definition of pagerank1 power method2}
\end{equation}
However, as it is formulated the PageRank algorithm will not output a meaningful vector. We will need to patch the procedure in various ways. 

\subsection{Patching the Algorithm}

It can be seen that if there are dangling nodes, pages that have no outlinks, then the power method will output the null vector. 
If we consider the following example:
\begin{figure}[!h]
\includegraphics[keepaspectratio=true,width=.33\linewidth]{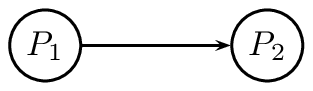}
\end{figure}

whose hyperlink matrix is:

\begin{equation*}
H = \left ( 
\begin{array}{cc}
\;0\; \;&0\; \\
\;1\; \;&0\; \\
\end{array} \right),
\end {equation*}

\noindent and start from $I_0 = (1,0)^t$ one ends up with $I = (0,0)^t $.

\noindent The first patch in the tinkering of the PageRank algorithm will be replacing the column corresponding to a dangling node with a column of all $1/n$ with n the number of nodes. This means that virtually every dangling node is linking to every single node in the web, including itself. This prevents the power method from giving the null vector. This way,  the disconnected graph becomes effectively connected at the price of giving a very low weight to the artificial bonds (added links).

\noindent The graph, with the addition of the extra links would look like:

\begin{figure}[!h]
\includegraphics[keepaspectratio=true,width=.483\linewidth]{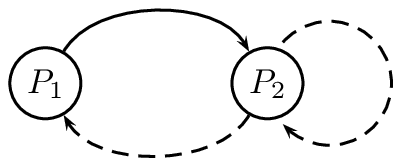}
\end{figure}
\noindent with a modified hyperlink matrix, $E$:

\begin{equation*}
E = \left ( 
\begin{array}{cc}
0&1/2\\
1&1/2\\
\end{array} \right)
\end {equation*} 
The matrix E that we obtain is, in general,   (column) stochastic, i.e. its columns all sum up to one. 
From the theory of stochastic matrices one knows that $1$ is always an eigenvalue. Furthermore,  the convergence of $I^k =  E I^{k-1}$ to $I$ depends on the second eigenvalue of $E$, $\lambda_2$. If it is smaller than $1$, then the power method will converge. In addition,  it is more rapid if $|\lambda_2 |$ is as close to $0$ as possible.

\noindent Let us consider the graph in fig.~\ref{fig:cyclicGraph}, with $E$ matrix:
\begin{figure}
\includegraphics[keepaspectratio=true,width=.33\linewidth]{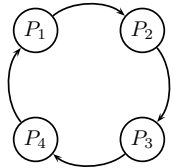}
\caption{
\label{fig:cyclicGraph}
(Color online) A graph whose matrix $E$'s second eigenvalue, $\lambda_2$, is one (see text in section~\ref{sect_II}).
}
\end{figure}
\begin{equation*}
E = \left ( 
\begin{array}{cccc}
0\;&\;0&\;0&\;1\\
1&0&0&0\\
0&1&0&0\\
0&0&1&0\\
\end{array} \right)
\end {equation*} 

\noindent One can see that the second eigenvalue of $E$, $|\lambda_2|$ is equal to one in this case, and  actually all the eigenvalues are on the circle of radius $1$ in the complex plane. 
If we compute $I$ with the power method starting from, say, $I_0 = (1,0,0,0)^t$ it will fail to converge. 

We will need to patch the algorithm again to ensure that $|\lambda_2| < 1$. In order to guarantee it,  we will require $E$ to be \emph{primitive}, i.e. that there is an integer $m$ such that $E^m$ contains all positive entries. The meaning of this assumption is that the graph is such that any page is connected by a path of at least $m$ links to any other.

\begin{figure}
\includegraphics[keepaspectratio=true,width=.33\linewidth]{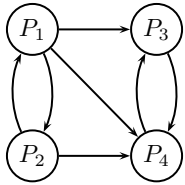}
\caption{
\label{fig:reducibleGraph}
(Color online) A graph that is not \emph{strongly connected}, or equivalently, whose matrix $E$'s is \emph{reducible} (see text in section~\ref{sect_II}).}
\end{figure}

Anticipating the interpretation of a diffusion phenomenon associated to searching the web,  we can interpret the requirement of $E$ to be primitive as the requirement of finding the walker with nonzero probability  on any site after a minimum time $m$.
Let us now consider the graph in fig.~\ref{fig:reducibleGraph}.
\noindent One can divide the graph in two subgraphs $\mathcal G_1$ and $\mathcal G_2$ . There are no links pointing from the subgraph $\mathcal G_2$, made of the nodes $3$ and $4$ to the first subgraph $\mathcal G_1$, made of nodes $1$ and $2$. If we write down the matrix $E$:
\begin{equation*}
E = \left ( 
\begin{array}{cccc}
0\;\;&\;\;1/2\;&\;\bf{0}\;&\;\bf{0}\;\\
1/3&0&\bf{0}&\bf{0}\\
1/3&0&0&1\\
1/3&1/2&1&0\\
\end{array} \right),
\end {equation*} 
we can see that there is a block that is zero, precisely the one that carries the information of the edges linking the nodes of $\mathcal G_2$ to the nodes of $\mathcal G_1$.

If we calculate $I$ with the power method starting from $I_0 = (1,0,0,0)^t$ we find: $I=(0,0,3/5,2/5)^t$. Yet, one is not  satisfied from giving an interpretation to the vector $I$ because the nodes from the first subgraph $\mathcal G_1$ have zero importance albeit being linked by other nodes. This is caused by  the  \emph{reducibility} of $E$ that causes a {\it drain} of importance from $\mathcal G_1$ to $\mathcal G_2$. In order to have a meaningful vector $I$, one that has all nonzero entries, one should demand that the matrix be \emph{irreducible}. A  necessary and   sufficient condition for it is that the graph is \emph{strongly connected}, i.e. that given two pages there is always a path connecting one to the other  (see~\cite{Meyer00}) chap. 8).

\subsubsection{The Patched Algorithm}

In order to implement all these patches, let us imagine a walk on the graph. With probability $\alpha$ the walker, which is equivalent to the surfer mentioned earlier, will follow the web with stochastic matrix $E$ and with probability $1-\alpha$, it  will jump to any page at random. The matrix of this process would be:
\begin{equation}
G := \alpha E + \frac{(1- \alpha )}{N} \mathbf{1},
\label{ eq: google matrix pagerank}
\end{equation}
 where $\mathbf{1}$ is a matrix with entries all set to $1$ and $N$ is the number of nodes. This matrix is known as the Google Matrix. Now, the matrix $G$ is irreducible because the matrix $\mathbf{1}$ is irreducible. Furthermore, it is also primitive since it has all positive entries. We have thus obtained a matrix that is both \emph{primitive} and \emph{irreducible}. This means it has a unique stationary vector that may be calculated with the power method. Furthermore, the result does not depend on the initial value $I_0$ because the underlying graph is strongly connected,  which is equivalent to the irreducibility of G, see~\cite{Meyer00}.

\noindent The parameter $\alpha$ is free and needs to be tuned. It is known~\cite{HaveliwalaKamvar03} that the second eigenvalue of $G$  , $\lambda_2$, is such that $|\lambda_2| \le\alpha $ , so one would choose $\alpha $ as close to zero as possible but in this way the structure of the web, described by $E$ would not be taken into account at all. Brin and page chose $\alpha = 0.85$ to optimize the calculation.
\subsubsection{Formulation as a Random Walk}

It is very appealing and useful to rethink the problem of assigning the importance of a page as the task of calculating the fraction of time a walker diffusing on the graph according to the stochastic process given by the Google matrix $G$. In fact we, can reformulate the Google PageRank as the algorithm that computes the fraction of the time the walker spends on each node by  defining the fraction on the $j^{th}$ page $T_j$ as:
\begin{equation}
T_i = \sum_{j} G_{ij} T_j.
\label{ eq: pagerank reformulation random walk}
\end{equation} 
Equivalently, one can say that the operational meaning of the PageRank algorithm is to give the probability to find the walker on the node $P_i$. Let us make it clearer by defining a set of random variables: $X^{(0)}, X^{(1)}, \ldots, X^{(n)}, \ldots  $ , one for every time  step. For each step,  the random variable can take on values in the set of nodes $\{ P_i \}$ of the web. 
We can recast Google PageRank in the language of a Markov Chain. Thus,  from eq.~(\ref{ eq: pagerank reformulation random walk}) written as:
\begin{equation}
{\rm Pr}(X^{(n+1)}=P_i) = \sum_{j} G_{ij} {\rm Pr}(X^{(n)}=P_j)
\label{ eq: pagerank reformulation markov chain}
\end{equation} 
and from the \emph{law of total probability}: 
\begin{equation}
\begin{matrix}
{\rm Pr}(X^{(n+1)}=P_i) = \\
\sum_{j} {\rm Pr}(X^{(n+1)}=
P_i | X^{(n)}=P_j) \, {\rm Pr}(X^{(n)}=P_j),
\label{ eq: pagerank reformulation markov chain 2 }
\end{matrix}
\end{equation} 
one can interpret the stochastic matrix $G$ as the conditional probability linking one time step to the other, i.e.:
\begin{equation}
G_{ij}  = {\rm Pr}(X^{(n+1)}=P_i | X^{(n)}=P_j) 
\label{ eq: pagerank reformulation markov chain G}.
\end{equation}
We will make use in the following of the latter interpretation of Google PageRank to devise methods to quantize it.

\section{Quantum PageRanks}
\label{sect_III}

 Quantum walks in their discrete time formulation were known already to Feynman~\cite{FeynmanHibbs65} and  since then, they were rediscovered many times~\cite{Aharonov93} and in contexts as different as quantum cellular automata~\cite{MayerA,MayerB} and the halting problem of the quantum Turing machine~\cite{Watrous,turing}.
For simplicity, let us discuss possible ways for quantizing a quantum walk taking place on the line. Later, we shall generalize it to an arbitrary graph.
 From now on, we shall make the notation lighter denoting each node (or page) $P_{i} $ simply by $i$ as shown in the following figure: 

\begin{figure}[!h]
\includegraphics[keepaspectratio=true,width=.93\linewidth]{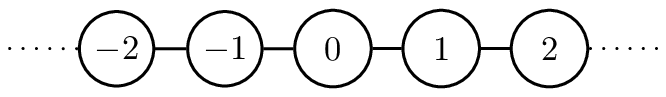}
\end{figure}

\noindent The naive way of quantizing this random walk would be to go from the index set $\{ i \,|\, i\in \mathbb{Z} \}$ to a Hilbert space of states ${\mathcal H } = {\rm span}\{ |\, i \,\rangle \,|\, i\in \mathbb{Z} \} $ as shown in figure~\ref{fig:naiveQuantumWalk}:

\begin{figure}
\includegraphics[keepaspectratio=true,width=.93\linewidth]{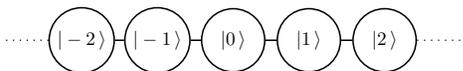}
\caption{
\label{fig:naiveQuantumWalk}
(Color online) The naive way to quantize a random walk on a line.
}
\end{figure}

\noindent Following the key idea outlined above, one could define the quantum importance of a page as:
\begin{equation}
I (P_i) = {\rm Pr}(X=P_i) :=   |\langle i |\, \psi \,\rangle|^2
\label{ eq: pagerank reformulation markov chain NAIVE }
\end{equation} 
where $ |\, \psi \,\rangle$ is the state of the system after it has diffused on the graph. However, this quantization procedure is not viable.  This is because the direct quantization of the time step evolution operator as
 \begin{equation}
 U =  \sqrt{p} \, | i+1 \rangle \langle  i | + \sqrt{1-p} \, | i-1 \rangle \langle i | 
\label{ eq: pagerank reformulation quantum walk NAIVE evolution }
\end{equation} 
is not unitary. Indeed,  while $\langle  0  | 2 \rangle = 0$ one has that $\langle  0 |  U^\dagger U| 2 \rangle \neq 0$ in the general case of $ p \in [ 0 , 1]$.

\noindent This difficulty can be overcome enlarging the Hilbert space. There are various ways to do it:
\begin{itemize}
\item adding a coin space ${\mathcal H_C}$ to each site on the quantum network. 
A coin space encodes the possibility to go left or right i.e. ${\mathcal H_C} = \rm{span} \{|L\rangle , |R\rangle\}$.
\item defining the Hilbert space as the space of (oriented) edges and treating the vertices as scattering centers. The resulting walk is called a Scattering Quantum Walk.
\item using Szegedy's~\cite{Szegedy04} procedure to quantize Markov chains.
\end{itemize} 
We will discuss the latter way for it will give us a valid quantization of Google's  PageRank that satisfies the properties  $1-4$ introduced in Sect.\ref{sect:intro}.
\subsubsection{Szegedy's Quantization of Markov Chains}
We have seen  that the Google matrix $G$ can be seen as the time step evolution operator of a Markov chain, or equivalently, of a discrete-time classical random walk: the two terms will be used interchangeably from now on.
 Szegedy put forward a general scheme to quantize a Markov chain. Let $G$ be a $N \times N$ stochastic matrix representing a Markov chain on an $N$-vertex graph. In order to introduce a discrete-time quantum walk on the same graph we use as the Hilbert space the span of all vectors representing the $N \times N$ (directed) edges  of the graphs i.e. ${ \mathcal H }=  {\rm span}\{ | i\rangle_1 | j \rangle_2 \, ,{\rm with}\, i , j \in N\times N\} = {\mathbb C}^{N} \otimes {\mathbb C}^{N}$. The order of the spaces in the tensor product is important here because we are dealing with a directed graph. We stress  this fact in the notation using subindices $1$ and $2$. 
 Let us define the vectors:
\begin{equation}
| \psi_j \rangle :=  |  j \rangle_1 \otimes  \sum_{k=1}^{N}  \sqrt{ G_{kj} } \,| k \rangle_2,
\label{ eq: Sgezedy vector 1}
\end{equation} 
that is a superposition of the vectors representing the edges outgoing from the $j^{th}$ vertex. The weights are given by the (square root of the) stochastic matrix $G$.

One can easily verify that due to the stochasticity of $G$ the vectors $ | \psi_j \rangle \, \rm{for}\, j=1,2,\ldots , N$ are normalized. The operator 
\begin{equation}
\Pi := \sum_{j=1 }^{N} | \psi_j \rangle  \langle  \psi_j |,
\end{equation}
 
is then a projector onto the subspace generated by the  vectors $ | \psi_j \rangle \, \rm{for}\, j=1,2,\ldots , N$. With this, the step of the quantum walk is then given by 
\begin{equation} 
U := S (2 \Pi -\mathbb{1})
\label{ eq: Sgezedy unitary walk operator}
\end{equation} 
where $S$ is the swap operator i.e. 
\begin{equation} 
S=\sum_{j,k=1}^{N} | j, k \rangle \langle k, j |.
\end{equation} 
  
Before continuing let us point out that the time step operator is unitary: 
\begin{equation}
U U^\dagger = S (2 \Pi - 1) (2\Pi -1) S^\dagger = S (4 \Pi^2 - 2 \Pi - 2\Pi +1)S^\dagger = 1
\label{ eq: Sgezedy unitarity time step 1}
\end{equation} 
from the unitarity of $S$ and the fact that $\Pi$ is a projector and squares to itself. Analogously:
\begin{equation}
U^\dagger U=  (2 \Pi - 1) S^\dagger S (2\Pi -1) = 4 \Pi^2 - 2 \Pi - 2\Pi +1 = 1
\label{ eq: Sgezedy unitarity time step 2}
\end{equation} 
The time step is thus the effect of a coin flip followed by a swap operator. Let us look more closely at the coin flip operation: 
\begin{equation}
2 \Pi -\mathbb{1} = \sum_{j=1 }^{N} \left( 2\, | \psi_j \rangle  \langle  \psi_j | - \dfrac{1}{N} \mathbb 1 \right).
\label{ eq: Sgezedy coin flip}
\end{equation} 

The vectors $| \psi_j \rangle  $ contain the information of the directed links that connect the $j^{th}$ node to all its neighbors to which it is connected through the stochastic matrix $G$. The sum on all nodes of the operators $ 2\, | \psi_j \rangle  \langle  \psi_j | - \frac{1}{N} \mathbb 1 $ is nothing but a reflection around the subspace spanned by the vectors $| \psi_j \rangle  $ and has the effect of enhancing the amplitudes of the mentioned directed edges at the expenses of the others. Furthermore, the swap operator preserves the unitarity of the step.

\subsubsection{Solving the Eigenvalue Problem for the Walk Operator} 

The quantization procedure based on the unitary operator \eqref{ eq: Sgezedy unitary walk operator} allows us to get remarkable insight on the properties of the walk by a systematic analysis of the spectral properties of $U$. The spectrum of the quantized walk is related to the spectrum of the original stochastic matrix that in turn we have seen has a key role on the properties of the related classical random walk.

Let us define the following $N\times N$ matrix $D$ that will play an important role in relating the spectra of the classical and quantum walks by specifying its entries
as follows: 
\begin{equation}
D_{ij} := \sqrt{G_{ij} G_{ji} 	\,},
\label{ eq: Sgezedy D matrix}
\end{equation}  
where there is no sum over repeated indexes. 
Let us also define the following operator $A$ from the space of vertices, ${\mathbb C}^{N} $, to the space of edges, ${\mathbb C}^{N} \otimes {\mathbb C}^{N}$:
\begin{equation}
A := \sum_{j=1 }^{N}   | \psi_j \rangle  \langle  j |
\label{ eq: Sgezedy A operator}
\end{equation}  
It has the following properties, which are straightforward to prove:
\begin{enumerate}
\item $\quad A^\dagger A = {\mathbb 1}$
\item $\quad A A^\dagger  = \Pi$
\item $\quad  A^\dagger S A  = D$
\end{enumerate}
The matrix $D $ is symmetric by construction. The eigenvalue problem $D | \lambda \rangle = \lambda | \lambda \rangle$ can in principle be solved  yielding $N$ eigenvalues $\lambda$ with the associated eigenvectors $| \lambda \rangle$:
\begin{equation}
\sigma(D):=\{ \lambda, |\lambda\rangle \}.
\label{D_spectrum}
\end{equation}

Let us now consider the the following vectors out of them: $| {\tilde \lambda} \rangle :=  A |  \lambda \rangle $ on the space of the quantized Markov chain i.e. 
${\mathbb C}^{N }\otimes {\mathbb C}^{N}$. In order to obtain the spectrum  of $U$, we will first isolate an invariant subspace for $U$ and look for eigenvalues and eigenvectors in this space. We will then concentrate on the orthogonal complement of it. We will argue that the interesting part of the Hilbert space for the dynamics is the aforementioned invariant subspace. In order to identify the invariant subspace of the walk operator let us see the effect of $U$ on $ | {\tilde \lambda} \rangle$:
\begin{equation}
U |{\tilde \lambda } \rangle = S (2 A A^\dagger - \mathbb{1}) A  | \lambda  \rangle =  S |{\tilde \lambda } \rangle,
\label{ eq: Sgezedy spectral prop walk 1}
\end{equation}  
where we used the property $1$ and $2$. Let us see also its effect on $S | {\tilde \lambda} \rangle$:
\begin{equation}
U S |{\tilde \lambda } \rangle = S (2 A A^\dagger - \mathbb{1}) S A  | \lambda  \rangle =  (2 \lambda S - {\mathbb 1}) |{\tilde \lambda } \rangle,
\label{ eq: Sgezedy spectral prop walk 2}
\end{equation}  
using properties $2 \,{\rm and }\, \,3$ and the fact that the vectors $| \lambda \rangle$ are eigenvectors of $D$. From (\ref{ eq: Sgezedy spectral prop walk 1}) and (\ref{ eq: Sgezedy spectral prop walk 2}) we can deduce that the subspace 
\begin{equation}
{\cal I}_U:=\{  |{\tilde \lambda } \rangle ,  S |{\tilde \lambda } \rangle \},
\label{invariant_subspace}
\end{equation}
is invariant  under the walk operator $U$. It is thus sensible to solve the eigenvalue problem:
\begin{equation}
U| \mu\rangle = \mu | \mu\rangle,
\label{ eq: Sgezedy spectral problem walk U}
\end{equation}  
for the walk operator restricted to the invariant subspace \eqref{invariant_subspace}.
Following what we have said,  let us make an educated ansatz for the eigenstates of U:
\begin{equation}
| \mu\rangle = |{\tilde \lambda } \rangle -\mu S |{\tilde \lambda } \rangle.
\label{ eq: Sgezedy spectral prop walk ansatz}
\end{equation}  
We have that: 
\begin{equation}
U | \mu\rangle = \mu  |{\tilde \lambda } \rangle + (1 - 2 \mu \lambda ) S |{\tilde \lambda } \rangle,
\label{ eq: Sgezedy spectral prop walk der 1 }
\end{equation}  
thereby the condition for $|\mu\rangle $ to be eigenstate of $U$ is
\begin{equation}
-\mu^2 = (1 - 2 \mu \lambda ),
\label{ eq: Sgezedy spectral prop walk der 2}
\end{equation}  
which yields $\mu = \lambda \pm \sqrt{\lambda^2 -1} = \exp (\pm \text{i} \arccos \lambda )$.

We note also that the span of the vectors $ |{\tilde \lambda } \rangle $ coincides with the span of the vectors $ |\psi_j \rangle $. Indeed we have $\sum_\lambda  |{\tilde \lambda } \rangle  \langle {\tilde \lambda } |  = A \sum_\lambda  | \lambda  \rangle  \langle  \lambda  | A^\dagger = \Pi = \sum_j  | \psi_j\rangle  \langle  \psi_j | $ . 

\noindent To complete our analysis let us point out that on the orthogonal complement to the span of the vectors $|\psi_j \rangle$,  the action of the walk operator $U= S(2\Pi - {\mathbb 1})$ is just $-S$,
which has eigenvalues $\pm 1$. This is because $\Pi$ yield the null vector when applied to vectors belonging to this subspace.  We conclude that the spectrum of $U$ is the set
 \begin{equation}
\sigma(U):=\{ \pm 1 , \exp (\pm \text{i} \arccos \lambda )\},
\label{ eq: Sgezedy spectrum of U}
\end{equation}  
where $\lambda$ are the eigenvalues of $D$.
In the following we will need the quantum walk where two steps at a  time are performed, with operator $U^2$. We can advance
that the interesting subspace, where we have dynamics, is the span of the vectors $|\psi_j \rangle$ and $S |\psi_j \rangle$ where the walk operator acts nontrivially.

\noindent Furthermore, considering the two-step evolution operator,  one can see that:
\begin{equation}
U^2 = (2S\Pi S - {\mathbb 1})(2 \Pi  - {\mathbb 1}).
\label{ eq: Sgezedy form of U2}
\end{equation}  
Therefore under the two-steps operator $U^2$ the Hilbert space splits naturally into the subspaces ${ \mathcal H }_{dyn}=  {\rm span}\{ | \psi_j\rangle , S | \psi_j \rangle  \}$ where dynamics takes place and its orthogonal complement ${ \mathcal H }_{nodyn}= { \mathcal H }_{dyn}^\perp$  as can be seen from \eqref{ eq: Sgezedy form of U2} $U^2$ acts trivially  in such a way that, obviously, ${ \mathcal H }= { \mathcal H }_{dyn} \oplus { \mathcal H }_{nodyn}$. The dimension of ${ \mathcal H }_{dyn} $
 is at most $2N$ and, remembering that the dimension of ${ \mathcal H } =  {\mathbb C}^{N} \otimes {\mathbb C}^{N} $ is $N^2$ we can conclude that the spectrum of $U^2$ corresponding to  ${ \mathcal H }_{dyn}$ is composed by, at most, the $2N$ values
\begin{equation}
\{ \exp (\pm 2 i \arccos \lambda )\},
\label{ eq: Sgezedy spectrum of U2 n1}
\end{equation}  
and the rest of the spectrum,  corresponding to 
${ \mathcal H }_{nodyn}$ where $U^2$ acts trivially,  is  composed of  at least $N^2 - 2N$ $1$'s.

The analysis presented above will allow us to save computational resources when calculating the Quantum PageRank because of the presence of the invariant subspace ${ \mathcal H }_{dyn}$. A different type of problem is concerned with using a quantum computer to perform a quantum computation that might improve the efficiency of the classical PageRank algorithm. An adiabatic quantum computation can be set to to compute the classical PageRank vector. In this case the classical PageRank vector is encoded  in a quantum state and an adiabatic quantum computation is described in order to reach this state~\cite{Lidar11}.

\section{A Quantization of Google PageRank}
\label{sect_IV} 

In this section we define a valid Quantum PageRank and take advantage of the analysis presented above to provide an efficient algorithm for its calculation
as requested in Sect.\ref{sect:intro}.
A natural way to define a quantization of the importance of a node or page in the quantum network associated to a directed graph
is to exploit the connection with the Markov chain process in which the fictitious walker is now subjected to quantum superpositions of paths
throughout the quantum web. In this way, the instantaneous importance of a quantum web page, denoted as $I_q(P_i,m)$, is given by the probability of finding the walker in that page $P_i$ at the node $i$ of the network after $m$ time steps.
As we have said before, the Hilbert space of this quantum walk is the set of directed links of the graph, $\mathcal H = {\mathbb C}^{N} \otimes {\mathbb C}^{N} $ where the numbering of the vector spaces is meaningful  due to the directedness of the underlying graph  and the second space in the tensor product contains the information of where the directed link points to. It seems natural then to project onto a vector of this second space $| i \rangle_2  $ obtaining the quantum state  $| I_q(P_i, m) \rangle   $ that contains the superposition of the nodes that were linked to it. To quantify the importance one can then extract a positive number calculating the norm of $| I_q(P_i, m) \rangle $, and with this we obtain an instantaneous list of page ranks including quantum fluctuations of the network. Thus we expect its instantaneous value to oscillate in time as a result of the underlying coherent dynamics.

The method for  computing the instantaneous PageRank of  the page $P_i$ is to start from an initial vector $ \, | \psi (0) \rangle$ and to let it evolve according to the two-step evolution operator $U^2$ (in order to swap the directions of the edges an even number of times, thus preserving the graph's directedness).  Then, we need  to project onto $| i \rangle_2  $, and  finally to take the norm of the resulting quantum state:
\begin{equation}
I_q(P_i,m) =  \langle  \psi (0) | \,  {U^\dagger}^{2m} | i \rangle_2  \langle  i |  U^{2m} | \psi (0) \rangle.
\label{ eq: Quantization of PR importance}
\end{equation} 
In order to implement the full procedure, one starts from the stochastic matrix $G$ representing the classical Google search that we want to quantize, forms the matrix $D$ and obtains its spectrum $\sigma (D) = \{\lambda\}$. One then forms the states $ |{\tilde \lambda } \rangle = A  | \lambda \rangle $,  in terms of which the eigenvectors of the walk operator, $| \mu \rangle$, in the subspace where the dynamics takes place are written.

Using the spectral decomposition of $U$ one can then arrive at a closed analytical expression for our quantum instantaneous PageRank:
\begin{equation}
I_q(P_i, m) =  \left\| \sum_{\mu}  {\mu}^{2m}{}_2 \langle  i | \mu\rangle \langle \mu | \psi (0) \rangle \right\|^2,
\label{ eq: Quantization of PR importance decompos 2}
\end{equation} 

where $| \mu\rangle = |{\tilde \lambda } \rangle -\mu S |{\tilde \lambda } \rangle$, $\mu  = \exp (\pm i \arccos \lambda )$ and $ | \psi (0) \rangle$ is taken to lie on the \emph{dynamical} subspace. Note that due to the fact that $| i\rangle_2$ are a basis of ${\mathbb C }^N$, in other words $\sum_i  |  i \rangle_2\langle i | = \mathbb 1 $ , from ~(\ref{ eq: Quantization of PR importance}) one can see:
\begin{equation}
\begin{matrix}
\sum_i I_q(P_i,m) = \\  \quad = \langle  \psi (0) | \,  {U^\dagger}^{2m} | \sum_i |  i \rangle_2  \langle  i |  U^{2m} | \psi (0) \rangle =  1 \quad \forall m,
\label{ eq: Quantization of PR importance sum check}
\end{matrix}
\end{equation}

\noindent meaning that  in the quantum version of  Google PageRank we have that the normalization condition~(\ref{ eq: Quantization of PR importance sum check})  is preserved at all times allowing to interpret the quantity $I_q(P_i,m)$ as the instantaneous relative importance of the page $P_i$, thereby reproducing a basic sum rule that also holds in the classical domain. 

In order to integrate out the fluctuations arising from the coherent evolution we also introduce the \emph{average} importance of the page $i$ sitting on the $i^{th}$ node  $\langle I_q(P_i) \rangle$ as:
\begin{equation}
\langle I_q (P_i)\rangle:=\frac{1}{M} \sum_{m=0}^{M-1} \ I_q (P_i,m).
\label{mean_value}
\end{equation}
We also use its variance or standard mean deviation:
\begin{equation}
\Delta I_q (P_i):=\sqrt{\langle I_q^2 (P_i)\rangle-\langle I_q (P_i)\rangle^2},
\label{Variance_quantum_pageRank}
\end{equation}
as a measure of its fluctuations.

\noindent In order to obtain the Quantum PageRank values of the nodes of a digraph we apply the algorithm that comes out of the analysis presented above. Namely the steps one has to perform are:

\noindent {\em \bf Quantum PageRank Protocol}
\begin{description}
\item[Step 1/]  Write the Google matrix for the digraph ${\cal G}$.
\item[Step 2/] Write down the matrix $D$ (see eq.~(\ref{ eq: Sgezedy D matrix})) and calculate its eigenvalues and eigenvectors.
\item[Step 3/]  Find the eigenvectors and eigenvalues of the two-step quantum diffusion operator $U^2$ in the dynamical subspace $H_{dyn}$ 
(see section~\ref{sect_III} for the details).
\item[Step 4/] Extract the Quantum PageRank value in time \eqref{ eq: Quantization of PR importance decompos 2}, its mean \eqref{mean_value} and standard deviation 
\eqref{Variance_quantum_pageRank} starting from the initial condition $| \psi_0 \rangle = \frac{1}{\sqrt{N}} \sum_{i=1}^{N } | \psi_j\rangle $.
\end{description}

\section{Results: Simulations for Quantum PageRanks}
\label{sect_V} 

After developing a quantum version of the Google PageRank in the previous section, it is necessary  to apply it to specific networks and by means of simulations,
see how it behaves as compared with the classical algorithm of PageRank.

\begin{figure}
\includegraphics[keepaspectratio=true,width=.53\linewidth]{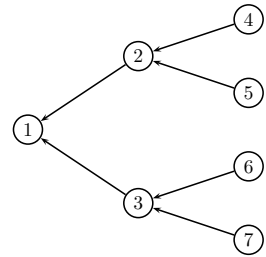}
\caption{
\label{fig:TreeThreeLevels} (Color online)The three level tree considered in the text to benchmark the Quantum PageRank (see text in section~\ref{sect_V}).
Each node represents a web page in an intranet with the root node being its home page.}
\end{figure}

\begin{figure}
\includegraphics[keepaspectratio=true,width=.53\linewidth]{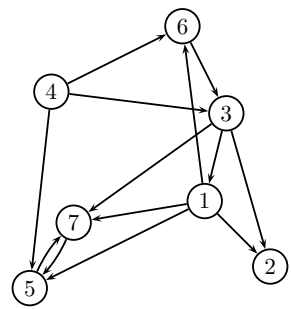}
\caption{
\label{fig:generalGraph}(Color online) The general graph with $7$ nodes considered in the text for benchmarking the Quantum PageRank (see text in section~\ref{sect_V}) .
Each node represents a web page in this directed quantum network.
}
\end{figure}

We have put to test our new quantum version of the PageRank algorithm in the case of a binary directed tree with 3 levels (see fig.~\ref{fig:TreeThreeLevels}) and a small albeit general, with no special property, directed graph (see fig.~\ref{fig:generalGraph}). We will describe the results in subsections~\ref{subsect_Tree Graphs} and~\ref{subsect_General_Graphs} respectively.

\subsection{Case Study 1: A Tree Graph}
\label{subsect_Tree Graphs}
In this subsection we will display the results obtained in the case of a tree graph (see fig.~\ref{fig:TreeThreeLevels}). 
This type of directed graph has a clear meaning in terms of a web network: it represents an intranet with the root node being the home page of a certain
website and its leaves representing internal web pages. This case of study has been extensively studied classically \cite{marijuan11}.
The quantum algorithm presented above is implemented numerically. 

\begin{figure}
\includegraphics[keepaspectratio=true,width=0.99\linewidth]{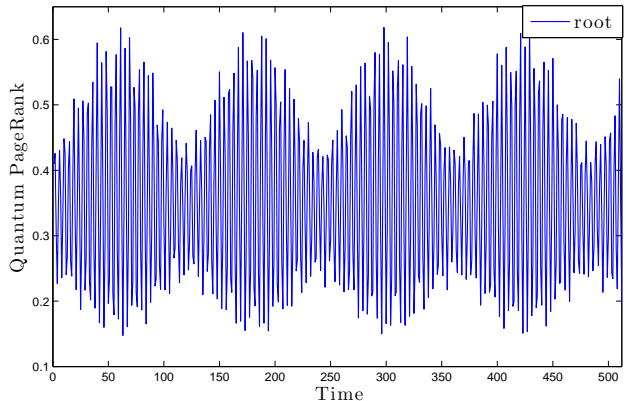}
\caption{\label{fig:qPR1TreeThreeLevels} (Color online) The evolution of the instantaneous Quantum PageRank $I_q$ with time for the  root page (home page)  in the case of a directed binary tree in Fig.\ref{fig:TreeThreeLevels}.}
\end{figure}

\noindent The quantum PageRank of the root page clearly oscillates in time and attains values that are higher than the classical counterpart, as shown in Fig.~\ref{fig:qPR1TreeThreeLevels}. According to the properties studied in Sect.\ref{sect:intro}, our quantum PageRank is {\em  instantaneously outperforming}. 

It is rather remarkable that
a quantum version of the PageRank may have an enhancement of the importance in the home page with respect to its classical counterpart. This is achieved merely by quantum means,
without changing the connectivity of the original directed graph as has been proposed classically  \cite{marijuan11}.

\begin{figure}
\includegraphics[keepaspectratio=true,width=0.99\linewidth]{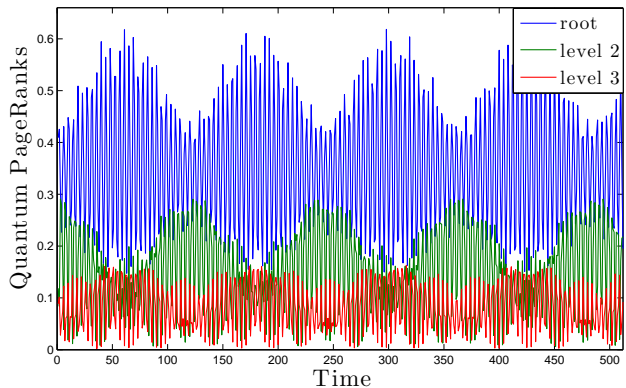}
\caption{\label{fig:realTimePlotTreeThreeLevelsQuantum} (Color online) The evolution with time of the instantaneous Quantum PageRank algorithm $I_q$ defined in Sect.\ref{sect_IV}
for web pages (nodes)  in the case of a directed binary tree graph in Fig.\ref{fig:TreeThreeLevels}. Only one page per level of the tree is displayed because pages that are in the same level have equal Quantum PageRank.}
\end{figure}

As for the quantum fluctuations present in the importance of the root page, it is important to emphasize that they remain bounded during the evolution as 
can be checked from Fig.~\ref{fig:qPR1TreeThreeLevels}. In addition, the classical value is always inside the range of the quantum fluctuations.
This feature is found to be true for all  pages in the directed binary tree network configuration (see Fig.\ref{fig:realTimePlotTreeThreeLevelsQuantum}).
A distinctive feature of the Quantum PageRank is that the hierarchy is not preserved at every time. Figure~\ref{fig:realTimePlotTreeThreeLevelsQuantum} clearly display the crossings of the instantaneous Quantum PageRanks.

\noindent In order to extract a fixed value for the relative importances of the pages, we compute  the mean value of the instantaneous quantum PageRank in time from eq.~\eqref{mean_value} and its variance from eq.~\eqref{Variance_quantum_pageRank}.

\begin{figure}
\includegraphics[keepaspectratio=true,width=0.99\linewidth]{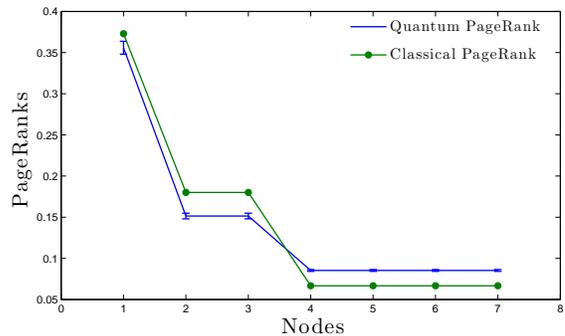}
\caption{\label{fig:Tree_Graph_3_levels_Classical_vs_Quantum} (Color online) Comparison of the hierarchies that result from the Classical and the Quantum PageRank in the case of the directed graph tree shown in Fig.\ref{fig:TreeThreeLevels}. Error bars in the quantum case are computed with the standard mean deviation
\eqref{mean_value},\eqref{Variance_quantum_pageRank}.}
\end{figure}

One can notice that the hierarchy of the pages is preserved on average and that the \emph{errors} i.e. the variances are negligible compared to the means. We can thus infer the pages' hierarchy as predicted by the classical algorithm from the Quantum PageRanks' averages as can be clearly seen in figure~\ref{fig:Tree_Graph_3_levels_Classical_vs_Quantum} for this particular case of directed binary tree graph.

\begin{table}[h] 

    \centering 
    
    \begin{tabular}{| c | c || c | c |} 
    \hline
    Level  & Classical PageRank  & Average Q-PageRank & Variance \\ 
    \hline\hline
    1& 0.37291 &0.355905 &0.0156461 \\
    \hline 
    2&0.18012&0.151437&0.0067747 \\ 
    \hline 
    3&0.06671&0.085305 &0.0022797 \\ 
    \hline 
    
    \end{tabular} 

    \caption{Comparison Classical and Averaged Q-PageRank in the case of the tree graph shown in Fig.~\ref{fig:TreeThreeLevels}. Given the symmetry of the tree only the values for each level are displayed. }

    \label{table:QPR_vs_CPR_tree_graph} 

\end{table}

Furthermore we can calculate a coarse grained evolution in time of the instantaneous Quantum PageRank. We divide its total evolution time of $M$ steps in $L$ equal segments made of an integer number $M/L$ steps. We calculate the mean in every segment:
\begin{equation}
{\bar I_q (P_i,n) }:=\frac{L}{ M} \sum_{m=n M/L}^{(n+1) M/L -1 } I_q (P_i,m) \; , n=0,\ldots , L-1.
\label{mean_coarse_grained}
\end{equation}

\begin{figure}
\includegraphics[keepaspectratio=true,width=0.99\linewidth]{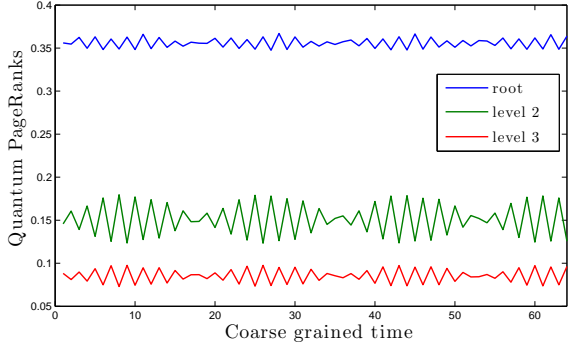}
\caption{\label{fig:meansPlotTreeThreeLevelsQuantum} (Color online)  The evolution in a coarse grained time (see text in section~\ref{sect_V}) of the Quantum PageRanks in the case of the directed graph tree shown in Fig.\ref{fig:TreeThreeLevels}. Again only one page per level is being displayed because pages that are in the same level have equal Quantum PageRank.}
\end{figure}

\noindent The result of integrating out the oscillations in a coarse grained time step is shown in figure~\ref{fig:meansPlotTreeThreeLevelsQuantum}. 
An interesting feature of this case is that the quantum PageRank oscillations shows a modulation with a clearly visible envelope for each node.
It strongly enforces the idea that the proposed algorithm has a solid grounding on the classical PageRank algorithm and it is a valid quantization of it.

\subsection{Case Study 2: A General Graph}
\label{subsect_General_Graphs}
We have performed a calculation of the quantum PageRank also in the case of a general directed graph with no particular symmetry (see fig.~\ref{fig:generalGraph}). 

\begin{figure}
\includegraphics[keepaspectratio=true,width=0.99\linewidth]{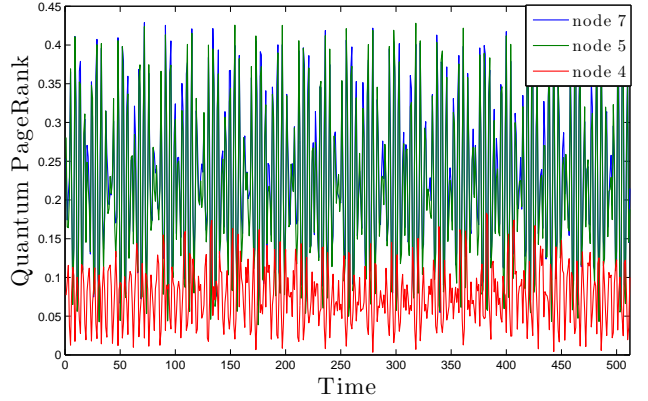}
\caption{\label{fig:realTimePlotGeneralGraph} (Color online)The evolution in time of the Quantum PageRank $I_q$ of  pages $7$, $5$ (that classically have the highest and nearly degenerate PageRank) and of page $4$ (that classically has the lowest PageRank) in the case of general graph shown in Fig.\ref{fig:generalGraph}.}
\end{figure}

\noindent The value of the instantaneous quantum PageRanks are found to display oscillations that are bounded and include the value found by the classical PageRank. The QPR of the node with the highest classical PageRank attains, at given times, values of the QPR that are higher than the classical counterpart (see fig.~\ref{fig:realTimePlotGeneralGraph}). As seen in the case of the binary tree treated above the Quantum PageRank is found to be {\em instantaneously outperforming} according to the definition given in section~\ref{sect:intro}.

\noindent The classical hierarchy is not preserved by the QPR at any given time (as is clearly shown by Fig.~\ref{fig:realTimePlotGeneralGraph}, see caption). We can notice crossings in the importance given by the instantaneous Quantum PageRank even between the pages that classically have highest and lowest PageRank. Furthermore, the nodes that have a very close classical PageRank are shown to have a very similar behaviour in time of their instantaneous Quantum PageRank. 

\begin{figure}
\includegraphics[keepaspectratio=true,width=0.99\linewidth]{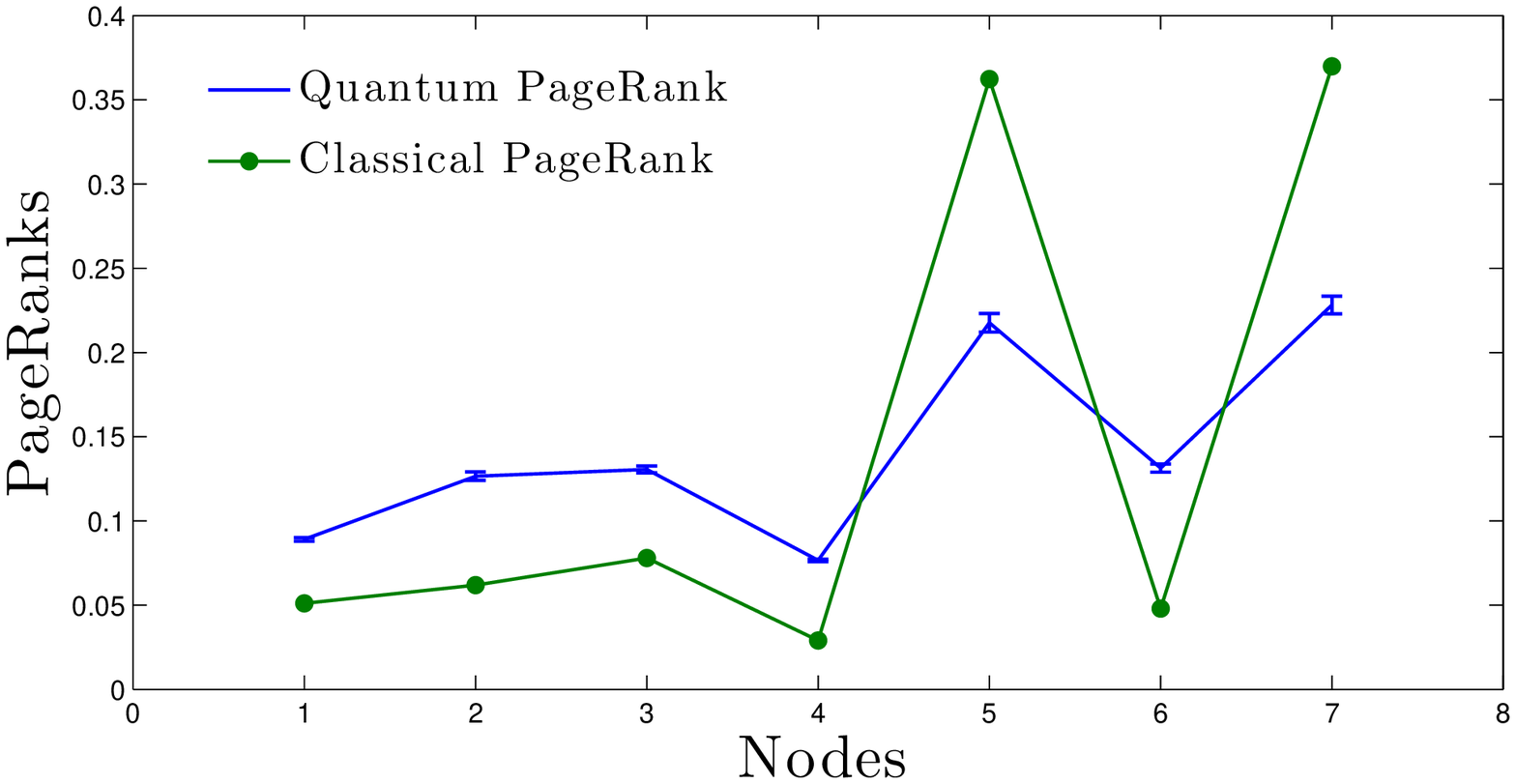}
\caption{\label{fig:withErrorsPlotGeneralGraph} (Color online)  Comparison of the hierarchies that result from the Classical and the Quantum PageRank in the case of the general graph shown in Fig.\ref{fig:generalGraph}. Error bars in the quantum case are computed with the standard mean deviation
\eqref{mean_value},\eqref{Variance_quantum_pageRank}.}
\end{figure}

Remarkably enough, we find that the  Quantum PageRanks' averages do not give us the same hierarchy as in the classical case (see Fig.~\ref{fig:withErrorsPlotGeneralGraph}).
Nevertheless, it is possible to clearly distinguish, within the error bar given by the variance,  which pages  have highest and lowest classical PageRank. 

\begin{figure}
\includegraphics[keepaspectratio=true,width=0.99\linewidth]{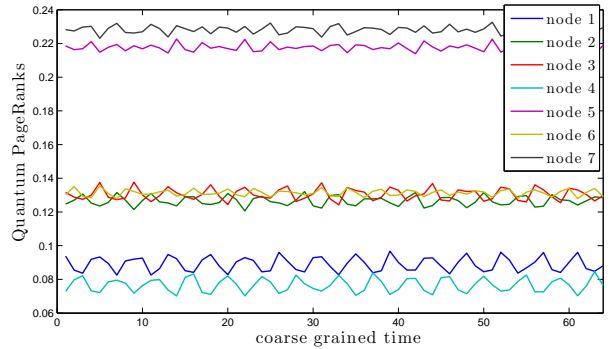}
\caption{\label{fig:meansPlotGeneralGraph} (Color online) The evolution in a coarse grained time (see text in section~\ref{sect_V}) of the Quantum PageRanks in the case of the general graph shown in Fig.\ref{fig:generalGraph}.}
\end{figure}

The analysis with a coarse graining in time of the instantaneous Quantum PageRank (see Fig.~\ref{fig:meansPlotGeneralGraph}) reinforces the conclusion that the classical PageRanks are still distillable in the case of pages with highest and lowest classical importances. 

\begin{table}[h] 

    \centering 
    \begin{tabular}{| c | c || c | c |} 
    \hline
    Node  & Classical PageRank  & Average Q-PageRank & Variance \\ 
    \hline\hline
    1&0.051019&0.089076&0.0021759 \\
    \hline 
    2&0.061860&0.126546&0.0050376 \\ 
    \hline 
    3&0.077924&0.130587 &0.0040337 \\ 
    \hline 
    4 & 0.028940 & 0.076586 & 0.0014675 \\ 
    \hline 
    5 & 0.362387 & 0.217691 & 0.0111097 \\ 
    \hline 
    6 & 0.047981 & 0.131345 & 0.0049477 \\
    \hline 
    7 & 0.369889 & 0.228169 & 0.010549 \\ 
    \hline 
    \end{tabular} 
      
    \caption{Comparison Classical and Averaged Q-PageRank in the case of the general graph shown in fig.~\ref{fig:generalGraph}.}

    \label{table:QPR_vs_CPR_gen_graph} 

\end{table}



\vspace{0.1cm}

\section{Conclusions}
\label{sect_conclusions}

In this paper we have proposed a notion of a class of protocols that qualify to be considered
a quantum version of the classical PageRank algorithm employed by the Google search engine (see Sect.\ref{sect:intro}).
In addition, we have constructed a step-by-step protocol explicitly in Sect.\ref{sect_IV} based on the quantization of 
Markov chains for directed graphs. This is a non-trivial problem since dealing with quantum versions of digraphs may produce
unitarity problems (see Sect.\ref{sect_III}). We have tested our quantum PageRank algorithm with two web networks in order to
gain insight into the specific behaviour of this protocol. One network is a binary tree graph representing an intranet with the root
being the home page of it. The other is a general directed graph with no specific structure.

From our numerical simulations we have found that our quantum PageRank has very interesting properties.
For the directed binary tree:

\noindent i/ The quantum PageRank for the root page is instantaneously outperforming with respect to the classical value.
This is a manifestation of the quantum fluctuations inherent to the quantum version of the algorithm and allows us to have
an enhancement of the importance of the root page without changing the topology of the original network.

\noindent ii/ The  instantaneous values of the quantum PageRanks for the nodes violate the hierarchy of the classical values.

\noindent iii/ The mean values of the quantum PageRanks including its standard deviation preserve the hierarchy of the classical values.

\noindent For the general directed graph:

\noindent i/ The quantum PageRank for the web page with the highest classical PageRank is some times higher than the classical
values obtained with the standard algorithm. This means that the quantum version of the PageRank 
is instantaneously outperforming with respect to the classical value.

\noindent ii/ The  instantaneous values of the quantum PageRanks for the nodes violate the hierarchy of the classical values.

\noindent iii/ Remarkably enough, there are pages with mean values, including its standard deviation, of their quantum PageRank
that violate the hierarchy of the classical values.

These properties are a clear manifestation that our proposal for a quantum version of the PageRank algorithm exhibits nontrivial features with respect to the Classical PageRank.

As the main purpose of our work is to devise a quantum PageRank algorithm by overcoming certain difficulties explained in the paper,
thus far we have dealt only with small networks representing different types of web. It would be very interesting to perform computations
with the quantum PageRank applied to very large networks with the properties exhibited by the complex structure of the real web 
\cite{small-world,scale-free,self-similar1,self-similar2,statistical1,statistical2,deterministic}.

\noindent An interesting issue is whether the classically first ranked page remains with the highest importance also at the quantum level. While we have shown that generically this is not the case for instantaneous values of the quantum PageRank (33) in Fig. 10 and Fig.13, they remain first-ranked with the time-average values of the quantum PageRank (35).  It remains open to see what happens for larger networks having very similar first-ranked nodes when the effect of quantum fluctuations are taken into account. This will depend on the topology of the lattice as well.

\begin{acknowledgments}
We thank the Spanish MICINN grant FIS2009-10061,
CAM research consortium QUITEMAD S2009-ESP-1594, European Commission
PICC: FP7 2007-2013, Grant No.~249958, UCM-BS grant GICC-910758.
\end{acknowledgments}


\end{document}